\documentclass[11pt]{article}
\usepackage{times}
\usepackage{amstex}
\newcommand{\ein}{1\!{\rm l}}
 
\newcommand{\CC}{\mathbb{C}}

\newcommand{\prob}{P}
\newcommand{\probl}{p}
\newcommand{\nn}{\nonumber}
\begin{document}
%
%
%
\title{\textbf{Bell's inequality and a strict assessment \\ 
of the concept of ``possession''}
\thanks{SNUTP 01-008}
}
\makeatletter
\author{Zae-young {\sc Ghim}\thanks{zyg@fire.snu.ac.kr}
 and Hwe Ik {\sc Zhang}\thanks{hizhang@phya.snu.ac.kr} \\
\textit{School of Physics, Seoul National University, Seoul 151-747}}
\date{\today}
\makeatother
\maketitle
\begin{abstract}
It is argued that the concept of ``physical quantities
possessed by the system'' is both redundant and 
inappropriate. 
We have examined two versions of the concept of
``possessed values'':
one identical with the observed values and the 
other non-identical. 
Assuming the existence of such ``physical 
quantities possessed by the system'' and
subjecting it to the framework of Bell theorem
under very generous  condition allowing
even certain form of ``nonlocality'', one can
still arrive at the proper Bell's inequality.
With the experimental falsification of Bell's inequality, 
we conclude that the concept of 
``possessed values'' finds no place 
in proper physical reasoning.

{\tt PACS Numbers:} 03.65.Bz, 01.70.+w 
\end{abstract}
\baselineskip 16pt

\section{Introduction}

Discussions on the interpretation of
quantum mechanics paved a way for a new discipline called
meta-dynamics which
deals with the foundation of the physical theories.
Experimental metaphysics, as baptized by A.\ Shimony, 
shows the fundamental aspects of the universal descriptional system
called dynamics. Especially Bell theorem\cite{bell1} verifies
that the fundamental issues themselves can be 
analyzed as those in usual physical problems,
especially in prediction-experiment manner.
In that sense, Bell theorem conforms a summit of
meta-dynamical investigations. However, its interpretation
is not unanimous in that its premises are not 
examined sufficiently enough to everybody's satisfaction.

It has widely been known that the concept of 
``possession by the system'' is problematique in 
quantum mechanics. 
But the conception of ``reality'',
which is apt to be identified with the existence of
values of physical quantities within the system,
is so strongly engraved in human imagination that we 
could not easily discard the concept of 
``possession of physical values by the system''.
The concept of nonlocality might be a 
convenient, if vague, concept to compromise to save the 
inconsistency between the quantum mechanics and 
this conception of reality.

Bell theorem gives a way of experimental approach 
to some important philosophical issues in physics, 
such as the concept of ``locality'' and ``reality''.
The experimental falsification\cite{expt} of Bell's
inequalities is regarded as a verification of
the failure of the theory based {\em both} on ``locality'' 
and on ``reality''.  Though
the latter is not the fully examined conception, especially
for quantum mechanics, the corroboration of 
quantum-mechanical prediction in experiment is 
usually interpreted as 
a vindication of the existence of ``nonlocality'' in Nature. 
However, both the notion of locality and reality must be defined more
precisely, since the proof, and accordingly the interpretation,
of Bell theorem might depend on them.

On the basis of such analysis,
we argue in this paper that the concept of ``physical quantities
possessed by the system'' is both redundant and 
inappropriate. It is redundant because it has no place
in the workings of dynamical theories other than its 
vague implication as the ``state of the system'',
which is well defined on its own stance, 
and it is inappropriate because it brings contradiction
in the light of Bell's inequality.
Two versions of the concept of ``possessed values''
are examined: one identical with the observed
values and the other non-identical. 
Assuming the existence of such ``physical 
quantities possessed by the system'' and
subjecting it to the framework of Bell theorem
under very generous  condition allowing
even certain form of ``nonlocality'', one can
still arrive at the proper Bell's inequality.
More specifically, the concept of nonlocality
here is divided into two categories, the ontological and
the epistemological one. The former is shown to be
perfectly permitted in the derivation, but the
latter is not in one version of the
``possessed values''. 
But since the epistemological nonlocality
defined here is highly unacceptable,
and the Bell's inequality is shown not to
be respected universally,
it is safe to conclude that the concept of 
``possessed values'' is in conflict with the
proper physical reasoning.

In next section we give a short review of Bell theorem, 
with a view to fixing the situation and summarizing
the received view on it.  It is followed by
a meticulous examination of the consequence of 
assuming that the system ``possess'' the values of the
physical quantities.  
(The appendix shows the detailed steps of the proofs.)
The final section discusses the meaning of
reality of physical quantities as well as  
elaboration of the concept of nonlocality in two
categories, viz., ontological and epistemological ones.
It leads to an alternative interpretation of Bell theorem.

\bigskip

\section{Received view on Bell theorem}

Consider a composite system made of two two-state subsystems,
such as two spin 1/2 particles or two polarized photons. 
Let \(\mathbf{A}\), \(\mathbf{B}\) be two physical quantities
restricted to each subsystem and 
\(\alpha\), \(\beta\) be their adjustable parameters 
representing the measurement setting,
such as the direction of the inhomogeneous magnetic field
in Stern-Gerlach apparatus or the angle of the analyzer for two-photon
cascade experiment.   
Denote by \(A_{\alpha}\) the presumed values 
of observables \(\mathbf{A}\) with the parameters of measuring
apparatus being \(\alpha\) and by \(a_\alpha\) the corresponding
outcome of the measurement.
For simplicity we abbreviate 
\(A_{\alpha}\), \(B_{\beta}\), \(A_{\alpha'}\), \(B_{\beta'}\)
by \(A\), \(B\), \(A'\), \(B'\) and the corresponding outcomes 
by \(a\), \(b\), \(a'\), \(b'\). 
We here are taking, deliberately, a most generous position
that the measured values \(a\), \(b\), \(a'\), \(b'\) 
may not be identical with the actually possessed values
\(A\), \(B\), \(A'\), \(B'\). 
(Throughout this paper, \(\mathbf{A}\) etc.\
denotes the physical quantity, 
\(A\) etc.\ the possessed value of \(\mathbf{A}\), 
\(a\) etc.\ the measured value.)

Bell theorem concerns with the coincidence rate defined by 
\begin{equation}
\xi(a,b)= \frac{\sum a b N(a,b)}{N_{\rm tot}} \ , 
\end{equation}
where \(N(a,b)\) means the number of the joint event
that the outcomes of two measuring apparatus
are given by \(a\) and \(b\), respectively.
\(N_{\rm tot}=\sum N(a,b)\) is the total number of
trials which is usually, but not necessarily, taken as infinity.
Other coincidence rates \(\xi(a,b')\), \(\xi(a',b)\), 
and \(\xi(a',b')\) are defined similarly.
\begin{quotation}
{\bf [Bell theorem]}\\
If we assume the locality and
the reality of physical quantities,
the particular combination of the coincidence rates defined by
\begin{equation*}
{\cal B}:= |\xi(a,b)-\xi(a,b')|
     +|\xi(a',b)+\xi(a',b')|
\end{equation*}
satisfies the following (Bell's) inequality 
\begin{equation*}
{\cal B} \le 2 \ .
\end{equation*}
\end{quotation}
The inequality above, called 
CHSH(Clauser-Horne-Shimony-Holt) inequality\cite{clauser}, is 
more general than the original inequality that
Bell derived. If one puts \(a'=b'=c\),
this inequality reduces to
the original one that
\begin{equation}
|\xi(a,b) -\xi(a,c)|
\le 1 + \xi(c,b) \ .
\end{equation}

According to Bell, the `vital assumption is that
the result \(b_\beta\) for particle 2 does not depend
on the setting \(\alpha\), of the magnet(i.e.\ measuring
apparatus) for particle 1, nor \(a_\alpha\) on 
\(\beta\),' which is usually 
taken as a requirement due to locality. 

In quantum mechanics, however, there exist a state for which 
\begin{equation}
{\cal B}_{\rm QM} > 2 \ .
\end{equation}
Therefore we can conclude that
`for at least one quantum mechanical state, the statistical 
predictions of quantum mechanics is incompatible with
separable predetermination' \textit{\`{a} la} Bell. 

With certain inspection regarding the basic presumptions
in the above theorem,
Bell\cite{bell1} concluded that
\begin{quotation}
\textbf{(Interpretation of Bell theorem 1)}  

In a theory in which 
parameters are added to quantum mechanics to
determine the results of individual measurements,
without changing the statistical predictions,
there must be a mechanism whereby the setting of
one measuring device can influence the reading
of another instrument, however remote. (\cite{bell}, p.\ 20)
\end{quotation}
One of the important concern of Bell was
the issue of (im)possibility of the hidden variable
theory, which can be defined to be the physical theory in which 
`the quantum mechanical states can be regarded as
ensemble of states further specified by additional
variables, such that given values of these variables
together with the state vector determine precisely
the results of individual measurement.' 
With this definition, one can say
\begin{quotation}
\textbf{(Interpretation of Bell theorem 2)} 
 
The quantum-mechanical result cannot be reproduced
by a hidden variable theory which is local. (\cite{bell}, p.\ 38)
\end{quotation}

Shimony declares a stronger version:
\begin{quotation}
\textbf{(Interpretation of Bell theorem 3)} 
 
No physical theory satisfying the specified 
independence conditions can agree in all
circumstances with the predictions of quantum
mechanics. (\cite{shimony}, p.\ 90)
\end{quotation}
For the case that the independence conditions
comes from the theory of relativity as locality
requirement, one can state
\begin{quotation}
\textbf{(Interpretation of Bell theorem 4)} 
 
No local physical theory can agree in all
circumstances with the predictions of quantum
mechanics. (\cite{shimony}, p.\ 91)
\end{quotation}

All of these ``received'' views on Bell theorem emphasize
that the assumption of locality makes main discrepancy
from quantum mechanical prediction. 
Logically, however,
Bell theorem has as its presupposition 
{\em both} the separability {\em and} the reality
of physical quantities, as well as other subsidiary
assumptions.  As Einstein 
expressed his firm belief that `the real factual
situation of the system \({\cal S}_2\) is independent
of what is done with the system \({\cal S}_1\),
which is spatially separated from the former',
it would be a better policy to examine
the other presuppositions before proceeding to find
fault with the assumption of locality. 
One of the main concern of present paper is to examine
the validity of these interpretations of
Bell theorem, on the basis of clearer analysis of
``locality'' and ``reality''.

\bigskip

\section{Derivations of Bell's inequality from the reality
of physical quantities}

The significance of Bell theorem stems not only from
its conflict with quantum mechanics, but also from
its derivation under certain conditions, 
irrespectively of any particular dynamics. 
Therefore, once Bell's inequality could be derived
under a clearly specified set of presumptions,
at least one of the presumptions can be blamed
to be ``wrong''. The main problem is how to
find out the most responsible presumptions to
be blamed.

The strategy we are taking in this paper is to
separate the presumably possessed values of physical
quantities from the actually measured values and
see clearly how they are involved in the process
of deriving Bell's inequality. By doing this,
we find that the most responsible, hence vital,
presumption underlying the derivation of Bell's
inequality is the very assumption that the
system possesses the values of physical quantities.
We will therefore examine very carefully
the possible consequence of the {\em assumption that
there exist the possessed values of physical 
quantities for the system, 
irrespective of quantum mechanics.} 
The values \(A\), \(B\), \(A'\), \(B'\) are presumably 
possessed by the system, and the actual measured 
values of them \(a\), \(b\), \(a'\), \(b'\) may or may not be 
different from them.

Note that while our consideration resembles the standard discussion
of Bell's inequality, the sole vital assumption here is that the
system possesses the values of physical quantities. 

\medskip

\subsection{Case 1}

We first consider the case where the measured values 
\(a\), \(b\), 
are numerically equal to \(A\), \(B\), 
which are assumed 
to take either \(+1\) or \(-1\). 
The number of possible combinations for 4 outcomes
\(a\), \(a'\), \(b\), \(b'\) is \(2^4\).

Now assume that the measured outcomes are the mere
uncovering of the values of physical quantities which
the system ``possesses''.

In this case, {\em whatever 
possible nonlocality is undergoing within the system},
the coincidence rate is given by 
\begin{equation}
\xi(a,b) 
= \frac{1}{N}\sum_{i=1}^{16} a_i b_i N_i  \ ,
\end{equation}
where  \(a_i\) and \(b_i\) denote possible value of 
the measured outcomes \(a\) and \(b\) respectively. 

From the positive-definiteness 
of probability, Bell's inequality is derived 
{\em \`{a} la} Wigner\cite{wigner}(See Appendix for detail.).
\begin{equation}
{\cal B}:= |\xi(a,b)-\xi(a,b')|
     +|\xi(a',b)+\xi(a',b')| \le 2
\end{equation}

  It should be noticed that the only assumption made 
here is that the system eventually came to possess 
four values \(A\), \(B\), \(A'\), \(B'\) without any regard how 
they come to be existent and the measured
outcomes are numerically equal to these possessed 
values. The possible correlations 
are fully allowed, and even the putative nonlocal 
influences between them are not excluded, since all 
we need is the outcomes which can be measured
explicitly.

\medskip

\subsection{Case 2} 

We now consider the more general case where the 
actual measured values \(a\), \(b\) of certain observables \(\textbf{A}\), 
 \(\textbf{B}\) can be unequal to the presumably possessed values 
\(A\), \(B\). Suppose that \(a\) and \(b\) can 
depend also on \(\alpha\), \(\beta\), the parameters 
representing the measuring apparatus (such as direction of the 
magnet), and on whatever hidden parameters \(\lambda\), 
and that the coincidence rate \(\xi(a,b)\) for the 
measurement \(a\), \(b\) is given by the expectation
value of the product of two possessed values
\begin{equation}
\xi(a,b) = \sum_\lambda A(\alpha, \lambda)B(\beta, \lambda )
\prob(A, B| \alpha ,\beta ,\lambda ) \ ,
\end{equation}
where 
\(\prob(A,B|\alpha ,\beta ,\lambda )\) is the joint probability 
for the event that 
the values \(A(\alpha,\lambda )\) and \(B(\beta,\lambda )\) 
to be observed as \(a\) and \(b\), respectively, given the
the parameters of apparatus and some hidden parameters. 
By this, we permit the possibility of \(\xi(a,b)\) 
depending on the setting of the measuring apparatus and also 
unknown hidden parameters \(\lambda\).
Without loss of generality we can choose the values of four 
physical quantities as 
lying in \([-1,1]\), which includes the dichotomic case. 

In order to derive Bell's inequality, it is necessary to 
assume the ``factorizability'' condition
\begin{equation}\label{factorize}
\prob(A,B|\alpha ,\beta ,\lambda ) = 
\prob_1(A|\alpha ,\lambda )\prob_2(B|\beta ,\lambda )
\probl(\lambda) \ ,
\end{equation}
where \(\prob_1(A|\alpha ,\lambda )\) and \(\prob_2(B|\beta ,\lambda )\)
are the probability distribution restricted to each subsystem
and \(\probl(\lambda)\) is the unknown, but arbitrary, distribution
for hidden parameters. 
The condition Eq.~(\ref{factorize}) 
requires both the value independence between \(A\) and 
\(B\), and the parameter independence between \(\alpha\)  
and \(\beta\). As it was noticed in Case 1, when we assume 
that the system possesses the four values \(A\),
\(B\), \(A'\), \(B'\), all the possible correlations between them 
are fully taken account, and even the possible 
nonlocal influences, if any, might have been taken 
account in the process of possessing such values. It 
is therefore quite natural to require the value 
independence between \(A\) and \(B\) in this context. 
[Otherwise, the measurement of the value \(A\) would 
lose any meaning because quantities other than \(A\) 
(not only \(B\) but any value in the world) would be 
involved in the process.] Also the assumption of 
parameter independence between \(\alpha\)  and \(\beta\) is 
rather benign. If we have to allow the parameter 
dependence, no meaningful experiment can be 
performed because any instrument located at any 
place in the universe may influence the values in a 
undetermined way, even if we disregard the 
celebrated `relativistic causality'.

From the assumption of factorizability, Eq.~(\ref{factorize}),
the coincidence rate is given by 
\begin{equation}
 \xi(a,b)= \sum_\lambda A(\alpha,\lambda) 
\prob_1(A|\alpha,\lambda ) B(\beta,\lambda)
\prob_2(B|\beta,\lambda ) \probl(\lambda) 
= \sum_\lambda \tilde{A}_\alpha(\lambda) 
\tilde{B}_\beta(\lambda) \probl(\lambda) 
\end{equation}
where \(\tilde{A}_\alpha(\lambda)=
A(\alpha,\lambda)\prob_1(A|\alpha,\lambda )\) and
\(\tilde{B}_\beta(\lambda)=B(\beta,\lambda)\prob_2(B|\beta,\lambda )\),
which do not exceed unity.
For these four numbers, it holds that(See Appendix for detail.)
\begin{equation}
{\cal Q}:= |\tilde{A}_{\alpha}\tilde{B}_{\beta} -  
 \tilde{A}_{\alpha}\tilde{B}_{\beta'}| 
+ | \tilde{A}_{\alpha'}\tilde{B}_{\beta} +  
    \tilde{A}_{\alpha'}\tilde{B}_{\beta'} |
\le 2 \ .
\end{equation}
%

From this inequality, we obtain
\begin{equation}
{\cal B}:=|\xi(a,b)-\xi(a,b')| + 
|\xi(a',b)+ \xi(a',b')| 
=\sum_\lambda {\cal Q} \probl(\lambda)
\le {\cal Q}_{\rm max} \sum_\lambda \probl(\lambda) 
\le 2 \ .
\end{equation}
where \({\cal Q}_{\rm max}\) means the maximum value of \({\cal Q}\).

Note that Eq.~(\ref{factorize}) can be compared with
Eq.~(10) of Ref.~\cite{bell81}. In the latter the
probability distribution \(\probl(\lambda)\) for 
the hidden parameters is absent. Though one can absorb
\(\probl(\lambda)\) mathematically into the definition of
\(\tilde{A}\) or \(\tilde{B}\), its physical meaning is
not unimportant. While we do not allow 
\(P_1(A|\alpha,\lambda)\) to be dependent on \(B\) or \(\beta\),
we don't have to care about the way how two
subsystems happen to possess the values \(A\) and \(B\)
respectively. This ignorance is generally 
expressed by a probability distribution for the
unknown parameters. In this sense \(\probl(\lambda)\)
shows that a sort of nonlocality is allowed, 
so long as it is \textit{not} observed at all and 
is prior to any measuring process. 

\medskip

\subsection{A variation of Case 2}

We examine here a variation of the previous case.
Instead of the factorizability condition, Eq.~(\ref{factorize}),
we can assume that the joint probability distributions for two physical 
quantities are not dependent on specific
parameter adjust for measuring apparatus, viz.,
\begin{equation}\label{thesame}
\prob(A, B| \alpha, \beta, \lambda)
= \prob(A, B| \alpha, \beta', \lambda) 
= \prob(A, B| \alpha', \beta, \lambda)
= \prob(A, B| \alpha', \beta', \lambda) \ .
\end{equation}
Invoking this assumption at the first step
of Eq.~(\ref{aineq4})(See Appendix for detail.),
we arrive at Bell's inequality. 
This is a rather strong assumption, since it means
that the joint probability of of the distant events
does not depend on the settings of the measuring apparatus
at all. In fact, it is stronger than our factorizability
requirement, Eq.~(\ref{factorize}).
In his own derivation of the inequality, Bell
employed the above assumption, Eq.~(\ref{thesame}) implicitly,
in addition to the explicit locality requirement, Eq.~(\ref{factorize}).

\bigskip

\section{Alternative interpretation of Bell theorem}

In the previous section, we considered the 
consequences of the assumption that
the system ``possesses'' the values of physical
quantities. For the Case 1, in which the 
measured values are assumed to be equal to 
the values which the system possesses, one can
derive Bell's inequality, fully permitting
any sort of nonlocality.
But, for the Case 2, in which the measured values
are not equal to the values presumably possessed
by the system, some added assumption is need
to arrive at the Bell's inequality.
In this case, one should invoke
either the `factorizability' condition,
Eq.~(\ref{factorize}), 
or the strong assumption of the identical
probability for the joint event of each
pair, Eq.~(\ref{thesame}).
But are these additional assumptions really matter?
What kind of locality does it demand or prohibit? 
Are the interpretations mentioned in
Sec.\ 2 still valid? 

In order to discuss these questions, 
it might perhaps be convenient to separate the 
concept of nonlocality into two parts: one the 
ontological one, and the other, the epistemological one. 
By the \textit{ontological nonlocality} we mean any possible 
nonlocal influences engendered in the process of 
determining the presumably existent values 
\(A\), \(B\), \(A'\), \(B'\) in the system, and 
by the \textit{epistemological nonlocality}, the 
presumable nonlocal influences occurring in the measuring process 
of obtaining the values  \(a\), \(b\), \(a'\), \(b'\) from the existent 
values  \(A\), \(B\), \(A'\), \(B'\).
 
The factorizability
condition, Eq.~(\ref{factorize}), for instance,
implies that we would have to allow the epistemological
nonlocality if we uphold the fact that 
there exist the values of physical quantities,
which are different from the measured values
of them but somehow possessed by the system.


 Using the above terminology, we summarize our 
conclusion as following: the measured values  \(a\), \(b\), \(a'\),
 \(b'\) corresponding to physical 
observables \(\mathbf{A}\), \(\mathbf{B}\) cannot be claimed 
to be existent in the system in general no matter 
what nonlocality is assumed, and the possessed values  
\(A\), \(B\), \(A'\), \(B'\) corresponding to 
\(\mathbf{A}\), \(\mathbf{B}\)
which are different from but somehow related to the measured values  
\(a\), \(b\), \(a'\), \(b'\) 
cannot be claimed to be existent in the system unless 
the highly unprobable epistemological nonlocality
is assumed. This shows that it is not legitimate for the 
system to possess the set of observed values. 
Moreover, it is also illegitimate
to have the set of presumed 
values related to the actual measured ones via 
possible parameters of the measuring apparatus or other hidden 
variables, unless we assume such a weird situation 
that any physical values or instrumental setups 
space-likely separated from the measuring activity 
can influence the measurement result.

On the other hand, the abandonment of the conception
of the possessed physical values does not cost at all
in physical theories.
The concept of possession of observable values by 
the system, or the concept of the existence of such 
values within the system does not have any 
significance even in classical physics. 
What is needed 
instead is the concept of the \textit{state} of the system.
In classical mechanics, for instance, the \textit{state}
of the system is represented by the pair of 
measured values, {\em not possessed values}, of 
position and momentum, which is used {\em only}
as the initial conditions for the prediction
of future events. It utterly does not concern
whether such values are possessed by the system
or not. In other words, the conception of
possessed physical values is fully redundant
in classical mechanics.

In quantum mechanics, however, the conception of 
possessed physical values is no longer redundant
but contradictory to the concept of the state.
As we already have seen, we have to accept 
an absurd conception like epistemological
nonlocality just to save this conception
without gaining any benefit in the
description of physical world.

Recently Ref.\cite{adenier} discussed a refutation of 
Bell theorem, maintaining that one cannot compare an
a local realistic inequality based on 
Eq.~(\ref{thesame}), which 
implies that `all four pairs of directions are
considered simultaneously relevant to each particle
pair', with the quantum mechanical prediction
derived from ``weakly objective interpretation''
which permits different joint probability distributions
for four pairs of observables. For `there is no way to derive a Bell
inequality' within the weakly objective interpretation. 
However, one can derive Bell's inequality with
a weaker presumption of the factorizability, Eq.~(\ref{factorize})
without assuming Eq.~(\ref{thesame}).
This shows very well that `within a strictly quantum mechanical 
framework' only the {\em state} of the system 
is considered and there is no place for the assumption
of ``possession of the values of physical quantities''.
We see that the conflict is {\em not} between local
hidden variable theory and quantum mechanics,
{\em but} between the assumption of reality of
physical quantities and the experiment.

Our consideration in this paper is deeply related to
the general investigations on the foundations of
dynamics in totality
and can be extended to an alternative interpretation
of quantum mechanics, as well as other dynamics, 
within a more systematic framework of 
meta-dynamics\cite{zhi94,ljw,meta4,book2k}.

\bigskip

\section*{Appendix: Detailed derivations of Bell inequality}
\numberwithin{equation}{section}
\renewcommand{\theequation}{A.\arabic{equation}}
\setcounter{equation}{0}
\renewcommand{\thesection}{\Alph{section}}
\setcounter{section}{1}

\subsection{Quantum-mechanical result}

As a typical example of the composite system of two-state 
systems which violates Bell's inequality, 
consider two spin 1/2 particles in a singlet state
\begin{equation}
|\Psi\rangle =\frac{1}{\sqrt{2}}\Big(
       |\uparrow\rangle_1 \otimes |\downarrow\rangle_2
       -|\downarrow\rangle_1 \otimes |\uparrow\rangle_2 \Big) \ .
\end{equation}
For this state, the coincidence rate is given by
\begin{equation}
\xi_{\rm QM}(a,b)=-\cos(\alpha-\beta)
\end{equation}
and 
\begin{equation}
{\cal B}_{\rm QM}
 = |\cos(\alpha-\beta)-\cos(\alpha-\beta')|+
       |\cos(\alpha'-\beta)+\cos(\alpha'-\beta')| \ .
\end{equation}
If we choose 
\begin{equation}
|\alpha-\beta'|=3\pi/4, \quad 
|\alpha-\beta|=|\alpha'-\beta|=|\alpha'-\beta'|=\pi/4 \ ,
\end{equation}
the violation of Bell's inequality is maximal:
\[ {\cal B}_{\rm QM}=2\sqrt{2}> 2 \ . \]

In the experiment\cite{aspect} by A.\ Aspect, P.\ Grangier and G.\ Roger,
for instance, the quantum-mechanical prediction is
\begin{equation}
{\cal B}_{\rm QM}= 2.70 \pm 0.05 \ ,
\end{equation}
with the incompleteness in the apparatus considered,
whereas the experimental result is
\begin{equation}
{\cal B}_{\rm exp}= 2.6970 \pm 0.015
\end{equation}
in great agreement with the quantum-mechanical prediction.

However, the above calculation has a loophole with the compatibility
of the associated observables.
Consider explicitly four self-adjoint operators acting on 
\(\frak{H}=\CC^2\otimes \CC^2\), such as
\begin{equation}
A= \vec \sigma\cdot \mathbf{n}_\alpha \otimes \ein\ , \quad
A'= \vec \sigma\cdot \mathbf{n}_{\alpha'}\otimes \ein \ , \quad
B=\ein \otimes \vec \sigma\cdot \mathbf{n}_\beta \ , \quad
B'=\ein \otimes \vec \sigma\cdot \mathbf{n}_{\beta'} \ .
\end{equation}
Note that while \([A, B]=[A, B']=[A',B]=[A',B']=0\),
it is no longer the case for the pairs \((A,A')\) and \((B,B')\). 
The above derivation demands that the system
should {\em possess} the values \((A, A', B, B')\) simultaneously, 
which is forbidden for quantum mechanics.

\subsection{Detailed proof for Case 1} 

Consider four physical quantities \(A_\alpha \), \(A_{\alpha'}\),
\(B_\beta\), \(B_{\beta'}\), which are assumed to be two-valued,
\(+1\) or \(-1\). Denote the outcomes of these physical quantities
by \(a\), \(a'\), \(b\), \(b'\), respectively. 
If these outcomes are just the values which the system ``possessed''
prior to measurement, the coincidence rate for pairs of physical 
quantities are plainly given by the simple expectation value
of the products of these.

The number of the possible combination of the products is 16. Let
the frequency for each case \(N_i\) (\(i=1,\cdots,16\)) and 
the corresponding probability \(p_i\). Then
for \(N\) measurements,   
\begin{equation}
p_i \equiv \frac{N_i}{N} \ , \quad N= N_1 + N_2 + \cdots + N_{16} \ , \quad
0\le p_i \le 1 \ , \quad i=1,\cdots,16
\end{equation}
It is convenient to have the following table: 
\begin{center}
\begin{tabular}{|c|cccc|cccc|}
\hline
{} & \(a\) & \(b\) & \(a'\) & \(b'\) 
& \({ab}\) & \({ab'}\) & \({a'b}\) & \({a'b'}\) \\
\hline 
1 & \(+\) & \(+\) & \(+\) & \(+\) & \(+\) & \(+\) & \(+\) & \(+\) \\ 
2 & \(+\) & \(+\) & \(+\) & \(-\) & \(+\) & \(-\) & \(+\) & \(-\) \\ 
3 & \(+\) & \(+\) & \(-\) & \(+\) & \(+\) & \(+\) & \(-\) & \(-\) \\ 
4 & \(+\) & \(+\) & \(-\) & \(-\) & \(+\) & \(-\) & \(-\) & \(+\) \\ 
5 & \(+\) & \(-\) & \(+\) & \(+\) & \(-\) & \(+\) & \(-\) & \(+\) \\ 
6 & \(+\) & \(-\) & \(+\) & \(-\) & \(-\) & \(-\) & \(-\) & \(-\) \\ 
7 & \(+\) & \(-\) & \(-\) & \(+\) & \(-\) & \(+\) & \(+\) & \(-\) \\ 
8 & \(+\) & \(-\) & \(-\) & \(-\) & \(-\) & \(-\) & \(+\) & \(+\) \\ 
 9 & \(-\) & \(+\) & \(+\) & \(+\) & \(-\) & \(-\) & \(+\) & \(+\) \\ 
10 & \(-\) & \(+\) & \(+\) & \(-\) & \(-\) & \(+\) & \(+\) & \(-\) \\ 
11 & \(-\) & \(+\) & \(-\) & \(+\) & \(-\) & \(-\) & \(-\) & \(-\) \\ 
12 & \(-\) & \(+\) & \(-\) & \(-\) & \(-\) & \(+\) & \(-\) & \(+\) \\ 
13 & \(-\) & \(-\) & \(+\) & \(+\) & \(+\) & \(-\) & \(-\) & \(+\) \\ 
14 & \(-\) & \(-\) & \(+\) & \(-\) & \(+\) & \(+\) & \(-\) & \(-\) \\ 
15 & \(-\) & \(-\) & \(-\) & \(+\) & \(+\) & \(-\) & \(+\) & \(-\) \\ 
16 & \(-\) & \(-\) & \(-\) & \(-\) & \(+\) & \(+\) & \(+\) & \(+\) \\
\hline 
\end{tabular}
\end{center}

Here the value of the column of \({ab}\) is just the product of 
two values of \(a\) and \(b\). Hence the value of 
\(\xi(a,b)\), for instance, is given by
\begin{align}
\xi(a,b)=p_1 &+ p_2 + p_3 + p_4 - p_5 - p_6 - p_7 - p_8 \nn \\
   &  - p_9 - p_{10} - p_{11} - p_{12} + p_{13} + p_{14} + p_{15} + p_{16} 
\end{align}
and similarly for \(\xi(a,b')\), \(\xi(a',b)\) 
and \(\xi(a',b')\).

Since the difference of two arbitrary positive number 
cannot be larger than their sum, we have
\begin{align}
|\xi(a,b)-\xi(a,b')| 
&=  2(p_2 + p_4 - p_5 - p_7 - p_{10} - p_{12} + p_{13} + p_{15} ) \nn \\
&\le  2(p_2 + p_4 + p_5 + p_7 + p_{10} + p_{12} + p_{13} + p_{15} )
\end{align}
and similarly
\begin{align}
|\xi(a',b)+\xi(a',b')| & = 
  2(p_1 - p_3 - p_6 + p_8 + p_{9} - p_{11} - p_{14} + p_{16} ) \nn \\
&\le  2(p_1 + p_3 + p_6 + p_8 + p_{9} + p_{11} + p_{14} + p_{16} ) \ .
\end{align}

Therefore we obtain
\begin{equation}
{\cal B}:= |\xi(a,b)-\xi(a,b')| 
    + |\xi(a',b)+\xi(a',b')| \le 
2 \sum_{i=1}^{16} p_i = 2  \ .
\end{equation}

\medskip

\subsection{Detailed proof of Case 2}
\label{proof2}
 
For arbitrary two real numbers \(y\), \(y'\) such that
\(|y|\le 1\), \(|y'|\le 1\),
\begin{equation}\label{aineq1}
|y\pm y'| \le 1\pm yy' \ ,
\end{equation}
because
\begin{equation}
|y\pm y'|^2 - (1\pm yy')^2 = y^2 (1-y'{}^2 ) + y'{}^2 - 1 
  \le 1-y'{}^2 + y'{}^2 - 1 = 0 \ ,
\end{equation}
where the last step results 
from \(|y|\le 1\).

Using Eq.~(\ref{aineq1}), for four real numbers 
\(x\), \(x'\), \(y\), \(y'\) whose absolute values are not larger than 1, 
\begin{align} \label{factora}
|xy-xy'| = |x|\cdot |y-y'| \le |x|(1-yy')\le 1-yy' \ , \\
|x'y+x'y'| = |x'|\cdot |y+y'| \le |x'|(1+yy')\le 1+yy'  \label{factorb}
\end{align}
and hence 
\begin{equation}
|xy-xy'|+|x'y+x'y'| \le (1-yy') + (1+yy') = 2  \ .
\end{equation}

For four numbers
\(\tilde{A}_\alpha\), \(\tilde{A}_{\alpha'}\), 
\(\tilde{B}_\beta\), \(\tilde{B}_{\beta'}\)
the magnitude of which do not exceed unity, one has 
\begin{equation}
{\cal Q}:= |\tilde{A}_{\alpha}\tilde{B}_{\beta} -  
 \tilde{A}_{\alpha}\tilde{B}_{\beta'}| 
+ | \tilde{A}_{\alpha'}\tilde{B}_{\beta} +  
    \tilde{A}_{\alpha'}\tilde{B}_{\beta'} |
\le 2 \ .
\end{equation}

Using the ``factorizability'' condition, Eq.~(\ref{factorize}), 
Bell's inequality is derived.
 
\medskip
  
\subsection{Detailed proof of a variation of Case 2}

We give another derivation of Bell's inequality, 
motivated by a recent proof\cite{accardi}. We indicated
the intermediate steps in order to make the assumptions
more explicit. 
As in Sec.\ \ref{proof2}, 
we have 
\begin{equation}\label{aineq2}
|A_{\alpha}B_{\beta} -  A_{\alpha}B_{\beta'}| 
+ | A_{\alpha'}B_{\beta} +  A_{\alpha'}B_{\beta'} |
 \le 2 \ .
\end{equation}
for four random variables 
\(A_{\alpha}\), \(A_{\alpha'}\), \(B_{\beta}\), \(B_{\beta'}\)
whose ranges lie in \( [-1, 1] \). 

Since for \( X\in [-1, 1] \),
\begin{equation}\label{aineq3}
|E(X)|=\Big|\sum_i X_i p_i \Big| \le \sum_i |X_i| p_i = E(|X|) \ ,
\end{equation}
we have
\begin{align}
|E( A_{\alpha}B_{\beta}) - E( &A_{\alpha}B_{\beta'}) |
+ |E( A_{\alpha'}B_{\beta}) + E( A_{\alpha'}B_{\beta'}) | \nn \\
& = |E( A_{\alpha}B_{\beta} -  A_{\alpha}B_{\beta'}) |
+ |E( A_{\alpha'}B_{\beta} +  A_{\alpha'}B_{\beta'}) | \nn \\
& \le  E( |A_{\alpha}B_{\beta} -  A_{\alpha}B_{\beta'} |)
+ E( | A_{\alpha'}B_{\beta} +  A_{\alpha'}B_{\beta'} |) \nn \\
& =  E( |A_{\alpha}B_{\beta} -  A_{\alpha}B_{\beta'}| 
+ | A_{\alpha'}B_{\beta} +  A_{\alpha'}B_{\beta'} |) \nn \\
& \le {\rm max} (|A_{\alpha}B_{\beta} -  A_{\alpha}B_{\beta'}| 
+ | A_{\alpha'}B_{\beta} +  A_{\alpha'}B_{\beta'} |) \nn \\
& \le 2 \label{aineq4} \ .
\end{align}
The equalities in the first and third line comes from the
linearity of expectation value. 
The inequality in the second
line is Eq.~(\ref{aineq3}) and the fourth line results from 
\begin{equation}\label{aineq5}
E(Y)\le \sum_i Y_{\rm max} p_i = Y_{\rm max} \sum_i p_i = Y_{\rm max} 
\end{equation}
where \(Y_{\rm max}\) is the maximum value which the random
variable \(Y\) can assume. We used the fact \(0\le p_i \le 1\).
The last line of Eq.~(\ref{aineq4}) is Eq.~(\ref{aineq2}).

The coincidence rate of two physical quantities is
given by the expectation value of the joint event,
\begin{equation}
\xi(a,b) = E( A_{\alpha} B_{\beta}) \ .
\end{equation}
Substituting this into Eq.~(\ref{aineq4}), we get
\begin{equation}
{\cal B}:= |\xi(a,b)-\xi(a,b')|
     +|\xi(a',b)+\xi(a',b')| \le 2 \ .
\end{equation}

Note that the crucial assumption in the 
first and third line of Eq.~(\ref{aineq4}) is that 
\begin{equation}
P(A_{\alpha}, B_{\beta}) =P(A_{\alpha}, B_{\beta'}) 
= P(A_{\alpha'}, B_{\beta}) =P(A_{\alpha'}, B_{\beta'}) \ , 
\end{equation}
for all four parameters \(\alpha\), \(\beta\), 
 \(\alpha'\), \(\beta'\).

\bigskip

\end{document}